\documentclass[12pt]{iopart}
\usepackage{graphicx}
\usepackage{color,soul}
\usepackage[framemethod=tikz]{mdframed}

\begin{document}

\title[Wide-Stopband Aperiodic Phononic Filters]{Wide-Stopband Aperiodic Phononic Filters}

\author{K Rostem$^{1,2}$, D T Chuss$^3$,
K L Denis$^2$ and E J Wollack$^2$}
\address{$ö1$ Department of Physics \& Astronomy, The Johns Hopkins University, 3400 N. Charles St., Baltimore, MD 21218}
\address{$ö2$ NASA Goddard Space Flight Center, 8800 Greenbelt Road, Greenbelt, MD 20771}
\address{$ö3$ Department of Physics, Villanova University, 800 E. Lancaster Avenue, Villanova, PA 19085}
\ead{krostem1@jhu.edu}
\vspace{10pt}

\begin{indented}
\item \today
\end{indented}

\begin{abstract}

We demonstrate that a phonon stopband can be synthesized from an aperiodic structure comprising a discrete set of phononic filter stages. Each element of the set has a dispersion relation that defines a complete bandgap when calculated under a Bloch boundary condition. Hence, the effective stopband width in an aperiodic phononic filter (PnF) may readily exceed that of a phononic crystal with a single lattice constant or a coherence scale. With simulations of multi-moded phononic waveguides, we discuss the effects of finite geometry and mode-converting junctions on the phonon transmission in PnFs. The principles described may be utilized to form a wide stopband in acoustic and surface wave media. Relative to the quantum of thermal conductance for a uniform mesoscopic beam, a PnF with a stopband covering 1.6-10.4 GHz is estimated to reduce the thermal conductance by an order of magnitude at 75mK.

\end{abstract}

%
\vspace{2pc}
\noindent{\it Keywords}: Phononic Meta-material, Phonon Filter, Thermal Conductance


\maketitle
 
%

\section{Introduction}

\begin{figure}[!b]
\centering
\includegraphics[width=8cm]{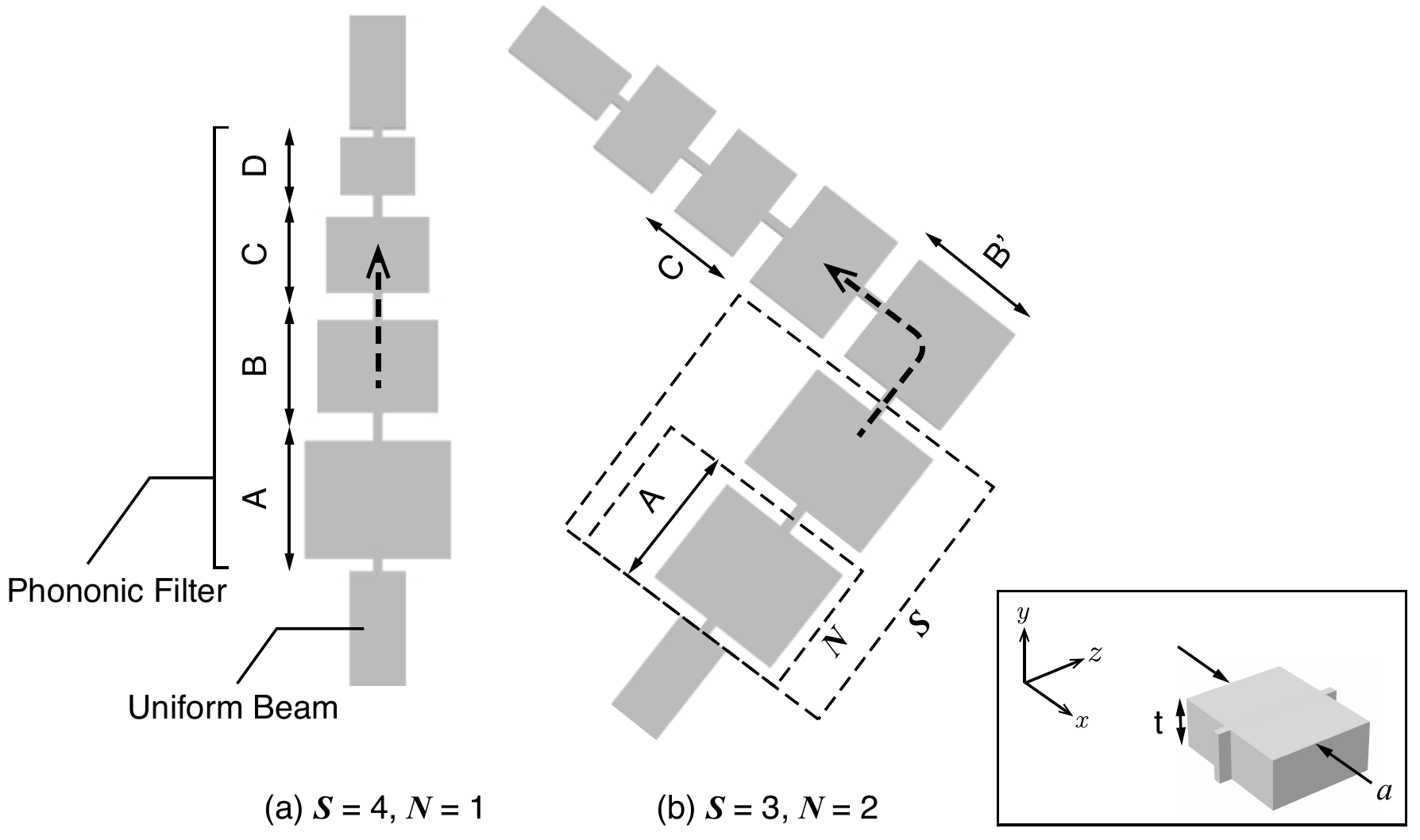}
\caption{\label{fig:phononics}~Example one-dimensional aperiodic phononic filters with wide stopbands. The individual filter stages employed are selected from geometries that result in complete phonon bandgaps when simulated in the periodic limit. The total stopband of the $S$$\times$$N$ stage filter extends from the lowest bandgap edge of cell A, to the highest bandgap edge of cell D. Each filter realization is utilized to explore a particular characteristic of a PnF, including a wide stopband in (a), and effect of mode conversion on low frequency phonons in (b). The inset shows a perspective view of a stepped-width unit cell. The thickness t is 300nm and constant across the planar phononic structures. The lengths A, B, B', C, and D, are 780, 645, 570, 550, and 400 nm respectively. The width $a$ is equal to the length in each stage. The uniform input/output beams in each filter are 300$\times$300 nm$^2$ in cross section, and matched to perfectly absorbing layers in finite-element computations. The dashed arrow indicates the transmission sense $+\vec k$ of the mechanical modes as defined by waveports in the model. }
\end{figure}

The ability to taylor the phonon dispersion curve in a meta-material phononic crystal (PnC) is ideal for managing heat transport. In addition to the phonon group velocity that can be engineered in a PnC to control the phonon flux~\cite{extremeG}, the existence of complete phonon bandgaps promises a new class of very low thermal conductance meta-materials~\cite{maldovan,Zen-Phononic}. In general, topology optimization is employed to maximize the bandgap width, $\Delta \omega$, arising from a unit cell arranged in an infinite periodic lattice. This approach is inherently limited by the fact that the lattice constant, $a$, fixes the length scale for resonant scattering and Bragg interference in the PnC. In addition, as the gap center frequency, $\omega_0$, is increased, the algorithms tend to generate progressively more complex shapes with feature size less than the lattice constant $a$~\cite{Osama,Dong,Kao}. Guided by these observations, we explore the filtering properties of a set of discrete coherence length scales combined to from a phononic filter (PnF) as illustrated in figure~\ref{fig:phononics}. In the periodic limit, the phonon dispersion of the geometry in each filter stage exhibits one or more complete phonon bandgaps. The bandgaps can be thought of as a filter pole, arising either from Bragg interference or resonance within the phononic structure. Thus, the effective stopband of a PnF is a consequence of several filter poles distributed across a wide bandwidth. 

In electromagnetism, the use of finite-sized structures as filtering elements is a concept that is well studied, for example, in the form of waveguide chokes, stepped impedance filters~\cite{collins}, and photonic structures~\cite{wide-PCJ}. A simple example is provided by a lumped capacitor and inductor combined to form a notch filter stopband. Adding additional lumped or distributed circuit elements to the filter can be used to adjust the response to achieve the desired stop-band specification. The work described here can be viewed in this context as the acoustic analog of a multi-mode transmission line structure. We refer to Brillouin~\cite{Brillouin} for a detailed discussion of how elastic bodies can be thought of or decomposed into elastic reactive elements. In contrast to electromagnetic single-mode waveguides, phonon waveguides are inherently multi-moded with at least four acoustic modes for a 1D beam (out-of-plane flexural, in-plane flexural, torsion, and compression)~\cite{Graff}, where D is the dimensionality of the reciprocal space. In practice, both the acoustic and higher order (optical) modes need to be considered. 


This work is motivated by the need for ultra-sensitive and fast cryogenic thermal sensors for future space-based far-infrared telescopes. The thermal-fluctuation-limited noise-equivalent-power of a bolometer is given by $\sqrt{4k_BT_0^2G}$, where $T_0$ is the operating temperature of the bolometer, and $G$ is the thermal conductance of the bolometer to the cryogenic environment~\cite{Irwin}. For state-of-the-art refrigeration systems capable of cooling a kilo-pixel detector array to 100 mK, reducing $G$ provides a path for reaching $\sim10^{-19}$ W/$\sqrt{\rm Hz}$ sensitivity without increasing cryogenic complexity. A wide stopband PnF can potentially limit the heat transport to the desired level in a compact structure less than 10 $\mu$m in lateral dimensions, an advantage given that a focal plane can be packed efficiently, and the filters will contribute negligible thermal mass to the sensor. 

In Section~\ref{sec:design}, we describe the framework of multi-moded 1D low-pass phononic filters. We solve the isotropic elastic wave equations using the finite-element method (see Appendix). In Section~\ref{sec:finite-geom}, we assess the effect of finite-geometry, and compute the transmission spectrum of the four lowest acoustic modes of a uniform beam through a phononic filter as illustrated in figure~\ref{fig:phononics}(a). In Section~\ref{sec:mode-conversion}, we describe the effects of a mode-converting junction (figure~\ref{fig:phononics}(b)) on the transmission of the compliant out-of-plane flexural and compression modes. In Section~\ref{sec:conductance}, we discuss application of the method in terms of thermal conductance and sensitivity of a bolometer thermally isolated with 1D PnFs.

\section{PnF Stopband\label{sec:design}}

\begin{figure}[!t]
\centering
\includegraphics[width=8cm]{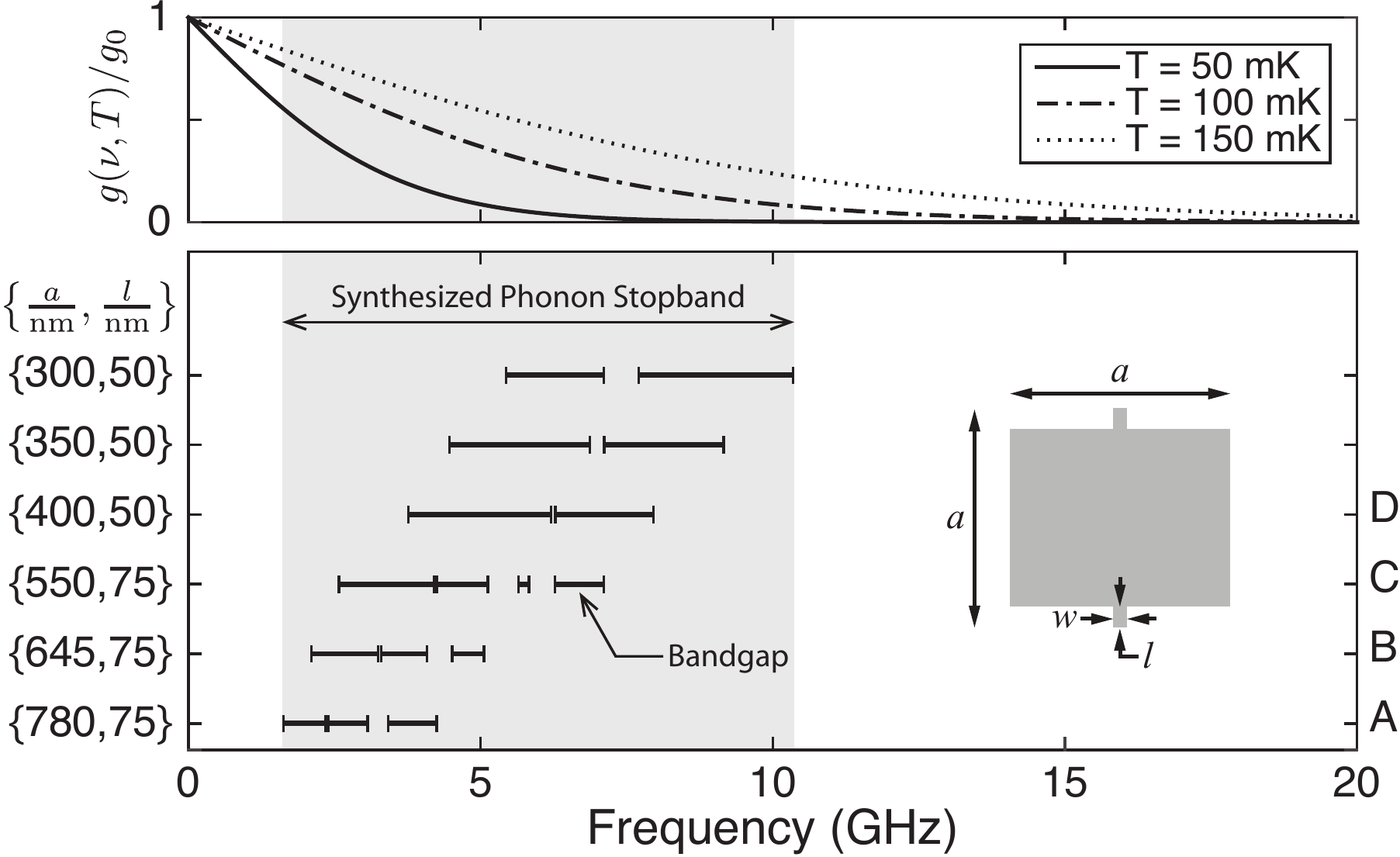}
\caption{\label{fig:parametric}~Complete bandgaps of a stepped phononic cell geometry calculated under Bloch boundary condition (lower panel). A cascaded aperiodic or quasi-periodic structure formed from a combination of these cells will exhibit a stopband covering 1.6 to 10.4 GHz (grey region). The cell dimensions are within the fabrication limits of state-of-the-art electron-beam lithography. A, B, C, and D refer to the cell geometries of the PnF shown in figure~\ref{fig:phononics}(a). The beam width $w$ is 50 nm in all cases. The stopband is compared to the effective bandwidth of a one-dimensional thermal source (upper panel). $g(\nu,T)/g_0$ represents the fractional cumulative thermal conductance of an acoustic phonon mode, where $g(\nu,T)=\int_\nu^\infty [dP(\nu,T)/dT] d\nu$, $P(\nu,T)$ is the one-dimensional phonon spectral density, and $g_0 \equiv g(0,T)$. }
\end{figure}

In a periodic phononic crystal, the bandgap center frequency, $\omega_0$, scales inversely with the size of the unit cell. This fundamental relationship that arises from the coherent properties of the crystal is the basis of the phononic filter approach. However, for a planar 1D structure with constant thickness, merely scaling the lattice constant, $a$, does not guarantee a precise shift in $\omega_0$. The frequency response is a function of the lattice constant to thickness ratio~\cite{Zen-Phononic}. Thus, we simulate phononic crystals with a Bloch boundary condition, select geometries that exhibit complete phonon bandgaps, and repeat the calculations as a function of $a$. Figure~\ref{fig:parametric} shows that for the material and unit cell geometry explored here, bandgaps across a continuous bandwidth from 1.6-10.4 GHz may be identified. Although other shapes could be used, the choice of a planar stepped-width unit cell reflects the relatively large complete phonon bandgaps that can be produced in a periodic crystal of this from. The step in width corresponds to a change in mechanical driving point impedance~\cite{Graff,stafford:localization}.

The unit cell geometries selected in figure~\ref{fig:parametric} can be cascaded to form a finite-sized phononic filter, as shown in figure~\ref{fig:phononics}. The filter would exhibit a phonon stopband that is synthesized from numerous poles distributed across a wide bandwidth. Poles arising from resonant scattering or Bragg interference may contribute to the response. We note that the stopband would formally transition to a complete phonon bandgap if a phononic filter is repeated as infinite periodic array. 

For the management of heat transport, the number of filter stages in a PnF can be treated as a free parameter when thermal isolation is the only objective, as is the case in thermo-electric generation. In applications where energy storage is also important, the rejection level and width of the stopband must be traded against increased thermal mass and parasitic effects, such as enhanced acoustic loss due to coupling to evanescent modes and fabrication imperfections. Thus, in the approach described here, $S\times N$ is minimized and the overlap of the bandgaps is optimized to reduce the thermal occupation of standing-wave modes where $d\omega/dk\simeq0$, and inter-band leakage.

The fractional bandwidth of the synthesized stopband in an aperiodic PnF can readily exceed 100\%. To evaluate the usefulness of a large rejection bandwidth, consider the broad phonon emission spectrum of a blackbody source through a 1D phononic channel, as shown in figure~\ref{fig:parametric}. For maximum effectiveness, the filter stopband is biased towards the Rayleigh-Jeans limit. However, even at 100 mK, 10\% of the thermal power is carried by phonons with frequencies greater than 10 GHz. Hence the typical 30\%~\cite{30percent} fractional bandwidth in a phononic crystal will have limited impact on the thermal conductance in a mesoscopic structure~\cite{Zen-Phononic}.

Above 1D, the PnF stopband should be placed over the peak frequency $\omega_{th}$ of the Planck distribution. In particular, the normalized bandwidth $\Delta \omega/\omega_{th}$ must be greater than unity to achieve substantial reduction in thermal conductance. To achieve this goal in a coherent structure such as a PnF, the cell size should span the appropriate range to enable a wide stopband. In a uniform beam with boundary-limited scattering~\cite{Rostem-2014}, the spectral density of the surface features needs to be sufficiently large in the waveband of interest to ensure diffusive phonon propagation. Fabrication tolerance ultimately determines the physical limit (coherent or diffuse) that can be successfully accessed to control heat flow. 

In a coherent phononic filter fabricated using electron-beam lithography, the degree of surface asperity is expected to induce negligible phonon decoherence. It has been shown that in electromagnetic meta-materials, decoherence becomes important when the scale of the surface roughness approaches the wavelength of the propagating electromagnetic mode~\cite{glushko}. A similar argument readily applies to elastic meta-material waveguides. Table~\ref{tbl:rough} summarizes the relevant roughness parameters in nano-meter features fabricated with electron-beam lithography~\cite{LER1}. The root-mean-square and correlation length in beam edge roughness are subdominant to the thermal wavelength at 100 mK by two orders of magnitude, $\lambda_{th}\sim 2\pi\hbar v/k_BT$, where $v$ is the typical speed of sound of out-of-plane flexural modes in a dielectric ($\sim$5000 m\,s$^{-1}$). Hence, cooling phononic devices to sub-Kelvin temperature offers the opportunity to fully test and measure the coherent long-wavelength meta-material properties of these structures at the mesoscopic limit~\cite{alegre,Zen-Phononic,Osman-FewModeBeam}.

\begin{table}[!b]
\caption{Typical surface roughness characteristics induced by electron-beam lithography~\cite{LER1}. For reference, the phonon thermal wavelength at 100 mK is shown. }
\begin{center}
\begin{tabular}{l|c}
\hline
\hline
Surface Roughness RMS & $\sigma_s\sim 5$ nm \\
Edge Roughness RMS & $\sigma_e\sim 5$ nm \\
Edge Roughness Correlation Length & $\xi_e\sim 100$ nm \\
Thermal Wavelength & $\lambda_{th}\sim 2000 $ nm  \\
\hline
\end{tabular}
\end{center}
\label{tbl:rough}
\end{table}

\section{Finite Geometry Effects\label{sec:finite-geom}}

The Bloch boundary condition is convenient for computing the elastic eigen-modes of an infinite crystal, however, a phononic crystal is truncated in a practical realization. The effect of finite geometry, or number of filter stages on the phonon transmission coefficient, is of therefore particular interest. 

\begin{figure}[!b]
\centering
\includegraphics[width=8cm]{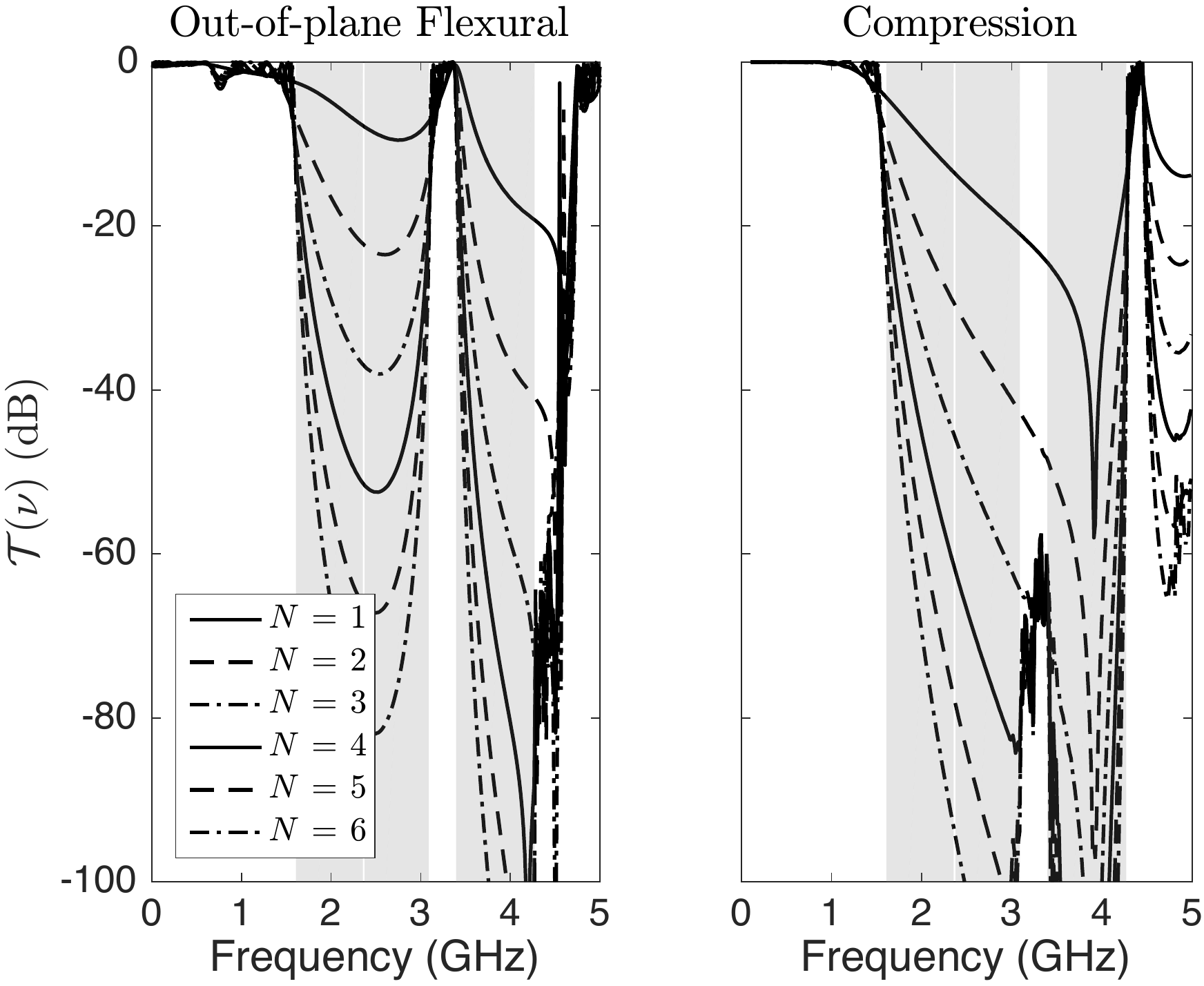}
\caption{\label{fig:N1M3}~Power transmission of the compliant acoustic modes through a quasi-periodic filter with $N$ filter stages ($S$ = 1). Referring to the unit cell geometry in the inset of figure~\ref{fig:parametric}, $a=780$ nm, $l=75$ nm, and $w=50$ nm. The grey region shows the bandgaps of the phononic crystal when calculated in the periodic limit. } 
\end{figure}

Figure~\ref{fig:N1M3} illustrates the typical change in transmission of the compliant out-of-plane flexural and compression modes of a uniform beam through a quasi-periodic $N$-stage filter ($S=1$). As $N$ is increased, the filter response asymptotically approaches that of an infinite array: the upper and lower stopband edges shift towards the ideal bandgap edges, and the transmission nulls deepen. Thus, as the crystal is truncated to a finite size, the complete phononic bandgap transitions to a stopband in which the phonon transmission of the propagating acoustic modes are reduced. When the wavelength is a significant fraction of the filter stage size, the minimum of the transmission coefficient in the bandgap regions scales exponentially with increasing $N$, as expected of evanescent decay in a coherent filter~\cite{collins,Savia,30percent}.

The required rejection of a phononic filter is application dependent, and typically determined by the signal strength, sensor noise and bandwidth. Three to five filter stages may be sufficient to provide the level of rejection needed for many applications~\cite{30percent}. As described in section \ref{sec:design}, in a practical implementation, increasing the number of filter stages should be traded against increasing thermal mass and decoherence arising from geometric imperfections.

\begin{figure}[!t]
\centering
\includegraphics[width=8cm]{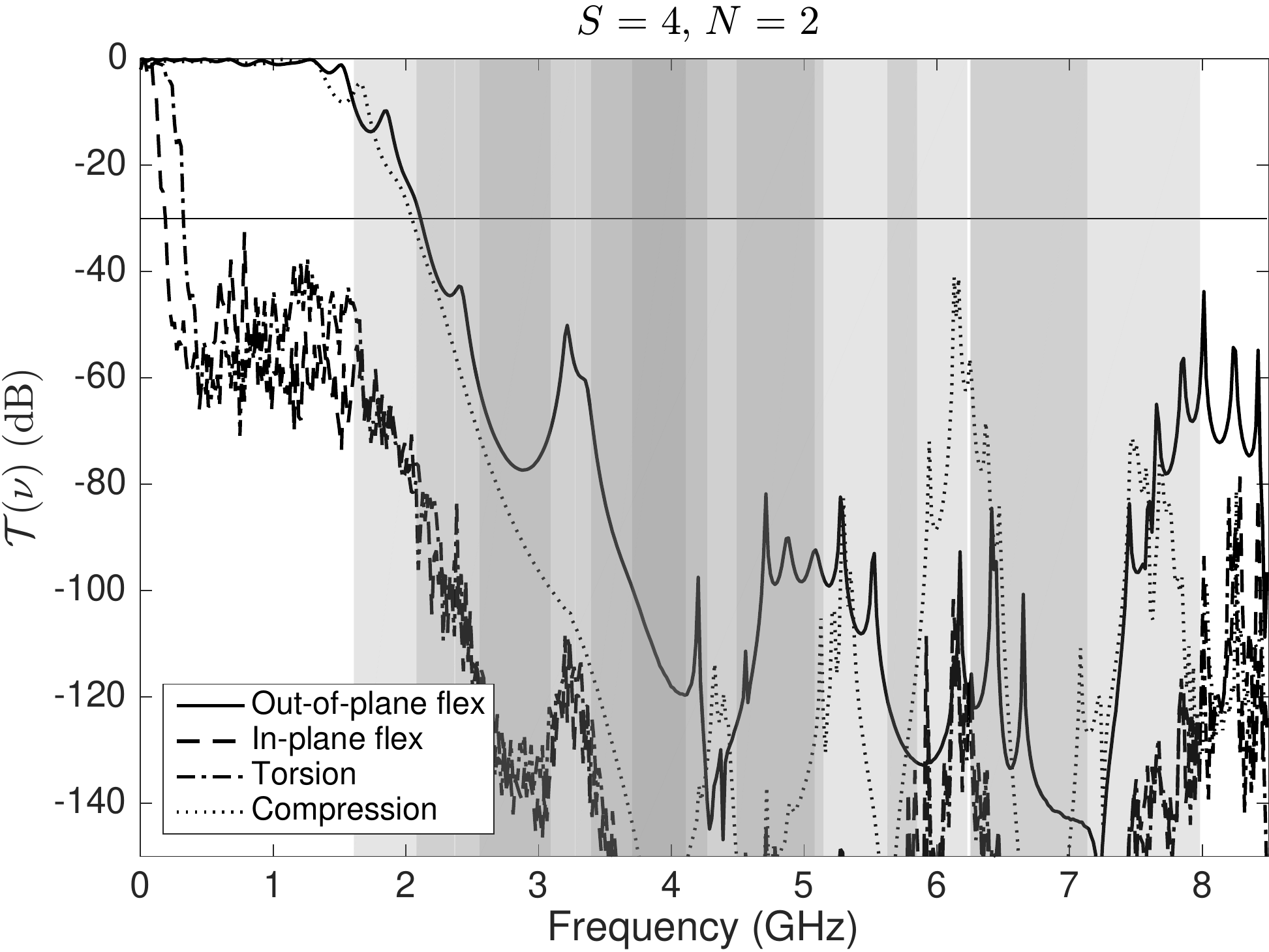}
\caption{\label{fig:N4M1}~Power transmission through a PnF with the geometry shown in figure~\ref{fig:phononics}(a). The grey region is formed by the overlapping bandgaps of each filter stage calculated in the periodic limit (see figure~\ref{fig:parametric}). The solid line at -30 dB is for reference. The frequency resolution in the simulation is 17 MHz.} 
\end{figure}

To illustrate a phonon stopband within a finite aperiodic geometry, we examine the filter shape shown in figure~\ref{fig:phononics}(a), with $S$=4 and $N$=2. $S$ and $N$ effectively define the stopband width and the level of rejection respectively. The transmission coefficient of the four acoustic modes is shown in figure~\ref{fig:N4M1}. The lower stopband edge for the out-of-plane flexural and compression modes occurs at 1.6 GHz, the lowest frequency of the first phonon bandgap in the filter calculated in the periodic limit (see figure~\ref{fig:parametric}). The in-plane flexural and torsion modes are strongly scattered due to the quadratic dependence of the mode impedance on the beam width~\cite{Cross,stafford:localization}. The out-of-plane flexural mode impedance also depends quadratically on thickness, however, because of the constant thickness of the geometry, these modes experience less rejection by the filter. The impedance of compression modes is linear in thickness and width. Thus out-of-plane flexural and compression modes are most susceptible to parasitic effects in the phononic filter geometry described here, although even at $N$=2, the transmission of the low-pass filter is below -30 dB from $\sim$2 to 8 GHz. We emphasize that the finite-element computations take into account the rich mode coupling between all the modes supported by the PnF, including evanescent modes.


\section{A Mode Converting Junction\label{sec:mode-conversion}}

Below the lowest stopband edge frequency of the PnFs, the compliant out-of-plane flexural and compression modes have near unity transmission. For the out-of-plane flexural mode, the transmission may be effectively limited if the thickness is varied across the filter geometry~\cite{Cross}. In applications where the device thickness is uniform, however, mode-conversion to the torsional and in-plane flexural modes may be employed to reduce the transmission of both compliant modes. 

\begin{figure}[!h]
\centering
\includegraphics[width=8cm]{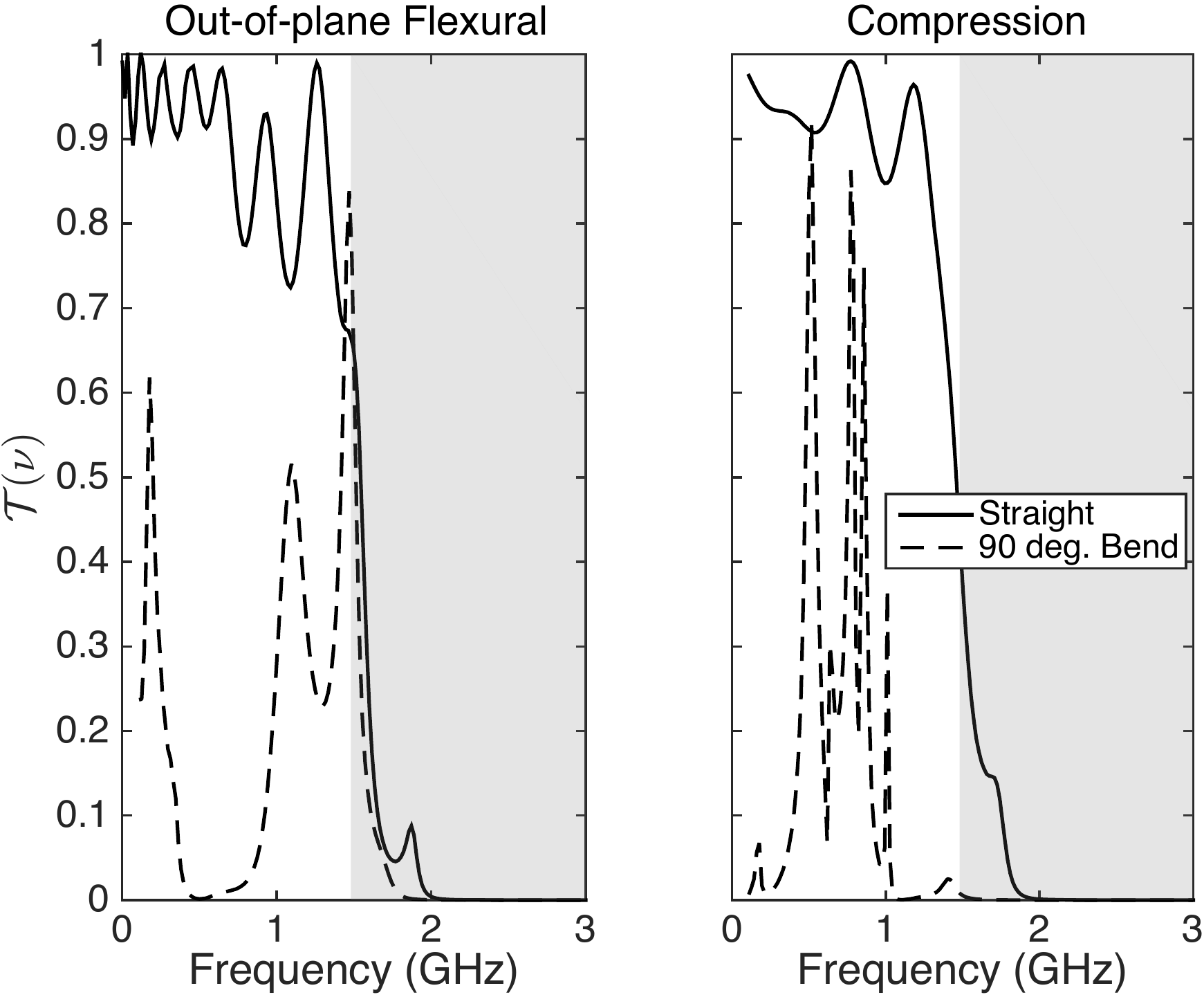}
\caption{\label{fig:OUTandCOMP}~Power transmission through the phononic filters shown in figure~\ref{fig:phononics}(a) and (b), solid and dashed lines respectively. $S$ = 3 and $N=2$ in both filters. The grey region is the bandgap of the first filter stage ($a=780$ nm) calculated in the periodic limit (see figure~\ref{fig:parametric}). }
\end{figure}

An example PnF with a mode-converting junction is shown in figure~\ref{fig:phononics}(b), where the 90$^\circ$ bend scatters out-of-plane flexural into torsional modes, and compression into in-plane flexural modes. By symmetry, the reverse scattering processes can also occur, however, the low transmission of the in-plane flexural and torsional modes across the PnF transfers negligible power. Relative to a PnF with the same filter stage dimensions but a straight profile, the 90$^\circ$ bend reduces the sub-Gigahertz average transmission by a factor of 5.5 for the out-of-plane flexural mode, and a factor of 4 for the compression mode, as shown in figure~\ref{fig:OUTandCOMP}. 

\section{\label{sec:conductance} Thermal Conductance of a PnF}

At 100 mK, the thermal wavelength exceeds the cross-sectional dimensions of the waveguides in figure~\ref{fig:phononics}. Thus, the thermal conductance of the uniform beam through the PnF can be estimated by incoherently summing over the contribution from the quantized 1D propagating modes, 
\begin{eqnarray}
G(T) = \frac{k_B^2 T}{2\pi \hbar} \sum_\alpha \int_{x_\alpha^{\rm c}}^{\infty} \mathcal{T}_\alpha(x) \frac{x^2 \exp(x)}{(\exp(x)-1)^2}dx,
\label{eqn:G}
\end{eqnarray}
where $x \equiv \hbar\omega/k_B T$, $\omega_\alpha^{\rm c} = x_\alpha^{\rm c} k_BT/\hbar$ is the cut-off angular frequency of mode $\alpha$, and $\mathcal{T}_\alpha$ is the power transmission coefficient through the PnF. For the four lowest acoustic modes, $\alpha=[1,4]$, $\omega_\alpha^{\rm c}=0$. These modes contribute to most of the thermal conductance when $T$$\ll$$\hbar\omega_1^{\rm c}/k_B$. Equation~\ref{eqn:G} is valid in the absence of acoustic loss, or thermalization, in a PnF. This approximation is reasonable given the expected level of decoherence induced by geometric imperfections in a practical device, as discussed in section~\ref{sec:design}. The upper bound on the total ballistic quantized thermal conductance, $G_Q$, can be evaluated with $\mathcal{T}_\alpha(x)$$\equiv$1, $\forall \alpha$. At 100 mK, the thermal conductance fraction carried by an acoustic mode extends far above the Rayleigh-Jeans equivalent bandwidth, $\pi k_BT/(12\hbar)=$ 3.4 GHz, as illustrated in figure~\ref{fig:parametric}. 

To establish a figure of merit, we use equation~\ref{eqn:G} to calculate the thermal conductance of a uniform beam with and without a PnF. Also of interest is the filtering performance of a PnF relative to a phononic crystal with a fixed lattice constant. For the latter, the cell geometry $\{a,l\}=\{780,75\}$ $\mu$m in figure~\ref{fig:parametric} is chosen as an ideal representative case with an infinite number of filter poles leading to unity transmission outside and zero transmission within the \emph{complete} bandgaps. The fractional widths are 20\% for the first, and $\sim$10\% for the second and third bandgaps respectively. 

Limited by the mesh size of the finite-element computations, we approximate the phonon transmission in the maximum possible synthesized stopband of the PnFs explored here (1.6-10.4 GHz, see figure~\ref{fig:parametric}) as follows. We compute the phonon transmission coefficient through the filter profiles in figure~\ref{fig:phononics} with $S=4$ and $N=2$ up to 8.5 GHz. The phonon transmission for every mode of the uniform beam in the 8.5-10.4 GHz range is set to -30 dB, a reasonable assumption given the expected rejection in each filter is an order of magnitude below this level. Above 10.4 GHz, the transmission is set to unity. The contribution of the optical waveguide modes of the 300$\times$300 nm$^2$ beam up to 30 GHz is included in all calculations.

\begin{figure}[!t]
\begin{center}
\includegraphics[width=8cm]{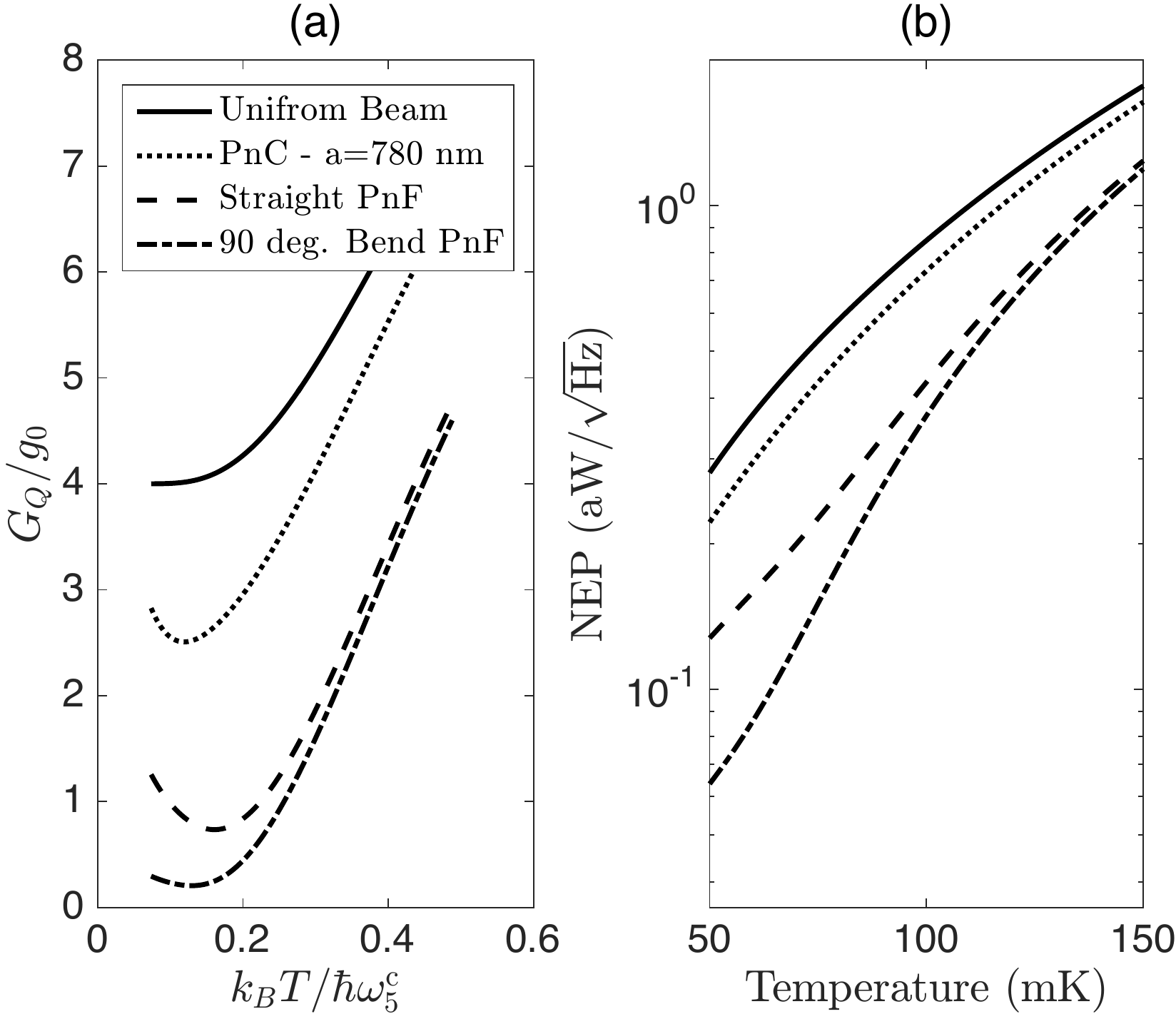}
\caption{\label{fig:G-NEP} ~(a) Thermal conductance of various 1D phononic waveguides, normalized to $g_0\equiv k_B^2\pi T/(6\hbar)$. The uniform beam cross-section is 300$\times$300 nm$^2$. All waveguides are matched to thermal baths at both ends. $\omega_5^{\rm c}/2\pi$=8.5 GHz is the cut-off frequency of the first waveguide mode of the uniform beam. $S=4$ and $N=2$ for the straight and bent phononic filters. (b) The thermal-fluctuation-limited noise-equivalent-power of a bolometer suspended by three filtered beams with conductance given in the left plot. An NEP $<4\times10^{-19}$ W/$\sqrt{\rm Hz}$ is achievable at 100mK. }
\end{center}
\end{figure}

Relative to the total quantum of conductance $G_Q$ for the uniform beam, an order of magnitude reduction in thermal conductance is achievable with a PnF below 75 mK, as shown in figure~\ref{fig:G-NEP}(a). In absolute terms, the estimated thermal conductance of a PnF with a 90$^\circ$ bend is 94 and 27 fW/K at 100 and 75 mK respectively. 

Depending on the application, a practical phononic filter may have one or both ends attached to a much wider membrane, essentially forming a discontinuous junction. At the low-frequency (long-wavelength) limit, the transmission of the acoustic modes tends to zero as a power law, $\nu^p$~\cite{Cross}. $p$ is 1/2 for the out-of-plane flexural mode, and 1 for the compression mode. Thus, the sub-Kelvin thermal conductance in figure~\ref{fig:G-NEP}(a) may be considered as an upper bound on the performance of a PnF as modeled with perfectly matched thermal baths at both ends of the filter. For the same reason, the noise-equivalent power of a phononic-isolated bolometer estimated in figure~\ref{fig:G-NEP}(b) is an upper bound on the sensitivity. 

\section{Conclusions}

We have explored the phonon filtering properties of aperiodic finite-sized multi-moded phononic waveguides. An aperiodic filter is composed of several cascading phononic stages of different size. The phononic geometry employed in each filter stage results in complete bandgaps when simulated in the periodic limit. Thus, the filter rejects acoustic and optical modes, and the synthesized phonon stopband may readily exceed 100\% in fractional bandwidth.

For the target application of bolometric sensors operating at 100 mK, an aperiodic phononic filter with a stopband from 1.6-10.4 GHz reduces the thermal conductance of a uniform beam by a factor of 5 at 100 mK, and by an order of magnitude at 75 mK. To achieve a similar conductance with boundary-limited scattering requires millimeter-long and micron-wide beams, in contrast to the compact phononic filter $<$10 $\mu$m in size. 

As envisaged, the aperiodic filter geometry has self-similarity. While the fractal dimension is not a unique descriptor of the filter profile, it may be useful in future studies when relating the geometric complexity to the filtering performance. As a result of the multi-moded nature of the problem, merely stating the size ratio of the largest to the smallest phononic shape in an aperiodic filter is not sufficient to describe its performance. Interestingly, if the geometry of the PnF changes according to a scaling law (e.g. logarithmically), the filter transmission and field pattern radiated into a higher dimensional elastic space would remain constant over the desired frequency bandwidth~\cite{Rumsey}. 

A multi-stage coherent filter can be applied in photonic~\cite{wide-PCJ}, acoustic, and surface wave media. A wide stopband phonon filter is particularly useful for thermoelectric generation, and for isolating high-Q mechanical resonators from thermal fluctuations in cryogenic optomechanical devices. 

\ack

We gratefully acknowledge financial support from the NASA Astrophysics Research and Analysis and Goddard Space Flight Center Internal Research and Development programs. 

\appendix
\section*{Appendix}

We assume the following elastic constants for a parent material that roughly corresponds to silicon in the ${<}100{>}$ direction~\cite{siliconE}: $E$ = 170 GPa, $\rho$ = 2330 kg m$^{-3}$, and $\nu$ = 0.28. All unconstrained boundaries are set as stress free. The finite-element computations inherently include coupling to all modes supported by a structure, including evanescent modes. 

Since the displacement and traction vector fields are complex valued, care must be taken to extract the phonon transmission coefficient for a given solution in a finite-element computation, i.e. the location of an excitation source matters relative to scattering regions such as step changes in geometry. Edge loads are used to excite the four acoustic modes. To investigate the level of interaction between the edge source and reflected fields, a uniform beam with perfectly matched layers at both ends (denoted by $AA$ for absorbing boundary conditions) is first simulated to ``calibrate" the response of the beam to a given load. The simulation is repeated with one end of the beam \emph{fixed} ($AF$). The reflection coefficient from the fixed boundary is evaluated from the complex displacement field, $r=U_{AF}/U_{AA} - 1$, and its deviation from unity is taken as a measure of the interaction strength between the source and the reflected fields. This procedure is generally known as a Thru-Reflect calibration~\cite{TRL}, and suggests an interaction strength of -30 dB at most between the source and reflected torsional acoustic mode power, and less than -50 dB for the other three acoustic modes.

\section*{References}
\bibliographystyle{unsrt}
\bibliography{References}

\end{document}